\documentclass[paper]{ieice}
\usepackage[dvips]{graphicx}
\include{epsf}

\SpecialSection{Nonlinear Theory and Its Applications}
\title{Plausible models for propagation of the SARS virus}
\authorlist{% fill arguments of \authorentry, otherwise error will be caused. 
\authorentry{Michael Small}{Non-member}{*}
\authorentry{Pengliang Shi}{Non-member}{*}
\authorentry{Chi~Kong Tse}{Non-member}{*}
}
\affiliate[*]{The authors are with the Department of Electronic and
Information Engineering of the Hong Kong Polytechnic University}
\received{2003}{1}{1}
\revised{2003}{1}{1}
\finalreceived{2003}{1}{1}

\begin{document}
\maketitle
\begin{summary}
Using daily infection data for Hong Kong we explore the validity of a
variety of models of disease propagation when applied to the SARS
epidemic. Surrogate data methods show that simple random models are
insufficient and that the standard epidemic susceptible-infected-removed 
model does not fully account for the underlying variability in the
observed data. As an alternative, we consider a more complex small world
network model and show that such a structure can be applied to reliably
produce simulations quantitative similar to the true data. The
small world network model not only captures the apparently random
fluctuation in the reported data, but can also reproduces mini-outbreaks
such as those caused by so-called ``super-spreaders'' and in the Hong
Kong housing estate Amoy Gardens.
\end{summary}

\section{SARS in the SAR}
%Severe Acute Respiratory Syndrome in the Special Administrative Region}

The redundantly named Severe Acute Respiratory Syndrome (SARS)
\cite{who03} appeared
in the Guangdong province of China in November 2002 and spread to the
Hong Kong SAR\footnote{That is, {\em Special Administrative Region} of
China, not to be confused with the virus.}. From Hong Kong, the virus
spread throughout the world: largely due to the airline (small world)
network hub in Hong Kong. The economic and social effects of the SARS
virus have been subject of much attention in the popular media \cite{scmp} and have
been widely reported. Less well reported, but no less unusual is the
effect this disease had on academic activities, including the cancellation
of the 2003 International Symposium on Nonlinear Theory and its
Applications. 

In this report we analyse the SARS daily infection data for Hong Kong
and test the validity of three distinct types of models: (i) stochastic
models generated from surrogate data; (ii) standard
susceptible-infected-removed (SIR) models; and (iii) small world network
models. We find that the small world network models are the only models
capable of reproducing the quantitative behaviour of the SARS epidemic in
Hong Kong. Moreover, these models also exhibit many features
characteristic of this epidemic: a small fraction of individuals show a
very great propensity to infect others (the ``super-spreaders''); and,
propagation of the virus within certain physical locations led to a
large number of infections (the outbreaks in Prince of Wales Hospital
and Amoy Gardens Housing Estate).

\begin{figure}[t]
\[\epsfxsize 65mm \epsfbox{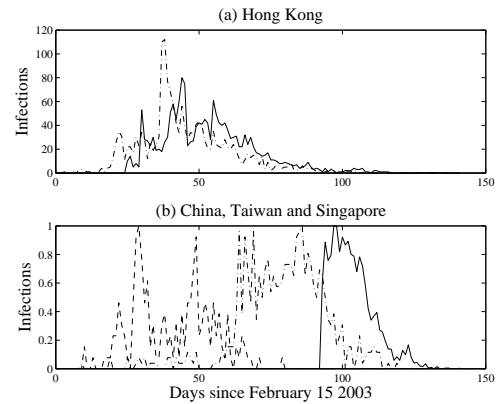}\]
\caption{{\bf Daily reported SARS infections.} The top plot shows the
daily number of SARS infections in Hong Kong. The lower plot are the
daily number of infections in Taiwan (dot-dashed), Singapore (dashed), and
the Chinese mainland (solid). For the purpose of comparison, the numbers
on the lower plot are normalised by the maximum daily infection
tally. The maximum daily infection tally for Taiwan, Singapore and China
was 26, 13 and 203. The data for Taiwan and Singapore has been revised
based on case information after the epidemic, this has not been possible
for the China data.}
\label{hksars}
\end{figure}

Figure \ref{hksars} shows the daily infection rate for SARS in Hong Kong
between 15 February 2003 and 15 June 2003. Compared to the data for
other territories (lower panel of Fig. \ref{hksars}), the data for Hong
Kong appears to be more complete (compared to Taiwan), more accurate
(compared to the Chinese mainland\footnote{In fact, various sources for
the Chinese mainland data showed widely divergent estimates. Furthermore,
subsequent analysis has shown that the original data significantly {\em
over}-estimated the effect of the virus.}) and over a longer time period
(compared to Singapore). The data for Hong Kong is shown as two time
series, the original daily number of infections, as recorded by the Hong
Kong Department of Health, and revised figures released after a
more detailed analysis of the true infection path \cite{doh03}. The revised
data offers a more accurate picture of the true daily number of
infections, but this data uses information not available at the time of
the outbreak and we therefore do not analyse that data in this report. 

For the current analysis, let $x_t$ denote the number of infections
reported in the Hong Kong media (based on statistics published by the
Hong Kong department of Health) for day $t$ and we start with
$x_0=0$. Hence $t$ is the number of days since 11 March 2003 (March 12
is the first day of recorded infections in Hong Kong, infections prior
to this date were only identified in the revised data). For
notational convenience we denote the entire time series as
$\{x_t\}_t=\{x_t\}_{t=0}^N=(x_0,x_1,x_2,\ldots,x_N)$.

The remainder of this paper describes the analysis of the original data
in Fig. \ref{hksars}(a) with three different model structures. In the
next section we describe the application of surrogate data techniques to
test the Hong Kong SARS time series against the hypothesis that
inter-day variability in the reported incidence of SARS is random. In
Section \ref{sir} we assess the accuracy of the SIR model and in Section
\ref{sw} we describe the small world network model.

\section{Random variability}
\label{surr}

The simplest model of disease propagation is that of a random walk such
that $x_0=x_N=0$ and $x_t\ge 0$ for $t=0,\ldots,N$. The duration of the
disease, $N$, is essentially a random variable. The
inter-day variation is random and, according to the Central Limit Theorem,
the disease will eventually die out.
  
\begin{figure}[t]
\[\epsfxsize 65mm \epsfbox{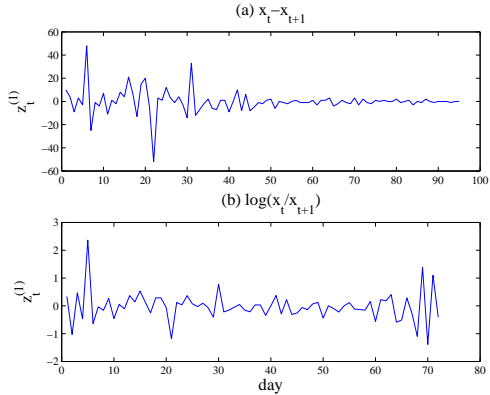}\]
\caption{{\bf Inter-day variability.} Variability between daily values
plotted as the difference $x_t-x_{t+1}$ (top plot), or log ratio
$\log\frac{x_t}{x_{t+1}}$ (lower plot).}
\label{diffdat}
\end{figure}

Our first task is to transform the data in Fig. \ref{hksars} to have
zero mean
 and be approximately independent and identically distributed
(i.i.d.). We do this by taking the difference of successive values
\[z^{(1)}_t=x_t-x_{t+1}\]
or taking the log-ratio of successive values
\[z^{(2)}_t=\log{\frac{x_t}{x_{t+1}}}.\]
Both $\{z^{(1)}_t\}_t$ and  $\{z^{(1)}_t\}_t$ are depicted in Fig.
\ref{diffdat}. We note that because there is an underlying pattern in
$\{|x_t|\}_t$ we expect that  $\{z^{(2)}_t\}_t$ will offer a time series
closer to i.i.d than  $\{z^{(1)}_t\}_t$.

We now apply the method of surrogate data \cite{jT92,casisurr} to
determine whether the transformed data in Fig. \ref{diffdat} is
statistically distinct from i.i.d. noise. Note that this is equivalent to
testing whether the raw time series $\{|x_t|\}_t$ is a random walk. 

To test for i.i.d. noise the method of surrogate data specifies that one
should generate an ensemble of {\em surrogate} data sets. Each surrogate
data set is a realisation of an i.i.d. process, but is otherwise ``like''
the original data. From this ensemble of surrogates one estimates some
statistical quantity and compares this distribution of statistic values
to the statistic value for the true data. If the true data has a
statistical value typical of the surrogates then the underlying
hypothesis (in this case, that the data is i.i.d.) may not be rejected. If
the data is atypical of the surrogate distribution then the hypothesis
may be rejected.

To generate suitable surrogates, we wish to destroy temporal correlation
but preserve the data's probability density function. The simplest way
to do this is by randomly shuffling either without \cite{jT92} or with
\cite{money} replacement. The choice of suitable test statistic is also
important \cite{pivotal_thing}. In our past experience we have advocated
correlation dimension \cite{pivotal_thing} or other dynamic invariants
\cite{casisurr}. However, for such extremely short time series this is
inappropriate. We attempted to use nonlinear complexity \cite{audiocomp} as a measure,
but found that this index exhibited extremely low sensitivity. We
instead settle on the normalised covariance between $x_t$ and $x_{t+1}$
\[\frac{<x_tx_{t+1}>}{<x_t^2>}.\]
One further advantage of this simple measure is that it is sensitive to
linear dependence: exactly the property we wish to
detect. Unfortunately, this also means that testing more complex linear
surrogate hypotheses will not be possible. Hence, as suggested by Takens
\cite{fT93} we prefer to test the model residuals against the
i.i.d. hypothesis. 

\begin{figure}[t]
\[\epsfxsize 65mm \epsfbox{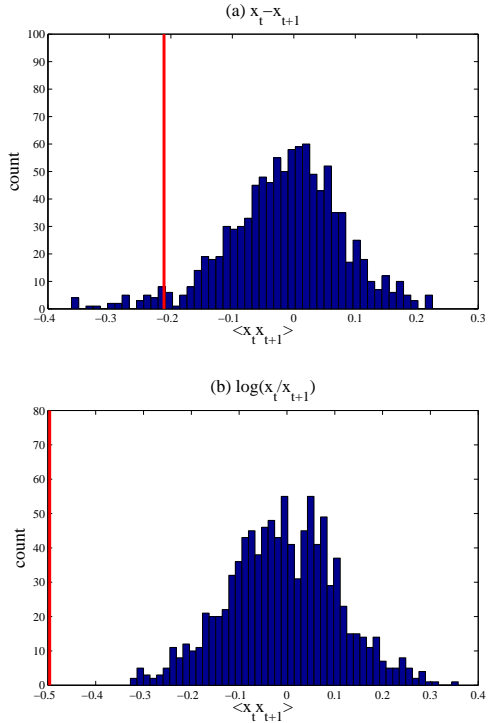}\]
\caption{{\bf Surrogate calculations.} The normalised covariance of $x_t$ and
$x_{t+1}$, is plotted for the two time series in Fig. \ref{diffdat},
along with $1000$ shuffled surrogates. In both cases the hypothesis that
the data in Fig. \ref{diffdat} is independent and identically
distributed noise is soundly rejected ($p>0.95$).}
\label{surrcalc}
\end{figure}

The results of Fig. \ref{surrcalc} clearly indicate that the data of
Fig. \ref{diffdat} is not consistent with an i.i.d. noise
source. Therefore the original data (Fig. \ref{hksars}) may not be
modelled as a simple random walk. 

It would be logical to continue this analysis with more sophisticated
surrogate algorithms \cite{jT92,cyclsurr}. However, for this short time
series results of more sophisticated algorithms are not
conclusive. Furthermore, as we have already noted, it would not be
possible to apply the covariance as a discriminating statistic in these
cases, as the surrogate algorithms explicitly preserve
autocorrelation. Instead, in the following section we consider a simple
deterministic model of disease propagation.

\section{SIR models}
\label{sir}

The standard Susceptible-Infected-Removed (SIR) model of disease
propagation has a long history and is considered in detail in any
standard text on mathematical biology \cite{jM93}. In difference
equation form, the relationship between the three populations at time
$t$ may be expressed as
\begin{eqnarray}
\label{sireqn}
\left(
\begin{array}{c}
S_{t+1}\\
I_{t+1}\\
R_{t+1}
\end{array}
\right) & = & 
\left(
\begin{array}{c}
S_t(1-rI_t)\\
(rS_t+(1-a))I_t\\
R_t+aI_t
\end{array}
\right)
\end{eqnarray}
where $S_t$, $I_t$, $R_t$ are the number (or proportion) of individual
that are susceptible to infection, infected and ``removed''
respectively. Not that, by removed, we mean individuals that have been
infected but are no longer infectious or susceptible. Such individuals
may be cured (and immune), quarantined, or dead. The parameters $r$ and
$a$ are the infection rate and the removal rate respectively.

\begin{figure}[t]
\[\epsfxsize 65mm \epsfbox{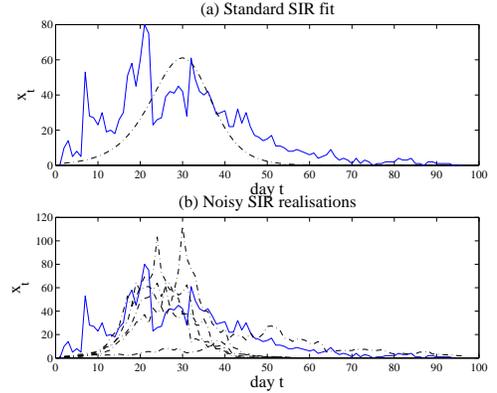}\]
\caption{{\bf SIR models of SARS.} The top plot shows the
daily number of SARS infections in Hong Kong and the best fit SIR model.
%($a=51.26$ and $r=7.05\times 10^{-6}$). 
The lower shows five simulations
from this model with multiplicative noise (described in the text) on the
daily number of new infections}
\label{sirfig}
\end{figure}

Using the SIR data we obtain a maximum likelihood estimate (i.e. minimum
least squares fit) of the parameters $a$ and $r$ for the Hong Kong SARS
time series. The time series in Fig. \ref{hksars} is the number of new
infections reported each day. This corresponds to the term $rS_tI_t$ in
Eqn. (\ref{sireqn}). The numerical fitting procedure often
estimates (incorrectly) a monotonically increasing of decreasing
function. To prevent these solutions, we add leading and trailing zeros
to the data prior to fitting. This helps to ensure that the optimal model
exhibits a single extremum --- not at either endpoint. An alternative to
fitting $rS_tI_t$ to the observed data would be to also consider the
daily fatality and recovery data for Hong Kong as an approximation for
$\Delta R_t$. One could then fit $\Delta I_t$ and $\Delta R_t$ directly
from the observed data.

In Fig. \ref{sirfig} (a) we see the time course of the best
model (Eqn. (\ref{sireqn})) fitted from this data. Clearly, this deterministic
model does not capture the random variation observed in the data. 

One way of considering this random variation is that the SIR model
provides an estimate of the rate of infection
(i.e. $I_{t+1}=(rS_t+(1-a))I_t$), but that this estimate is subject to random
fluctuations. It may be better therefore to perturb the number of new
infections by some unknown quantity $\epsilon_t$. The modified,
stochastic SIR equations then read
\begin{eqnarray}
\label{sireqn2}
\left(
\begin{array}{c}
S_{t+1}\\
I_{t+1}\\
R_{t+1}
\end{array}
\right) & = & 
\left(
\begin{array}{c}
S_t(1-rI_t)-\epsilon_t\\
(rS_t+(1-a))I_t+\epsilon_t\\
R_t+aI_t
\end{array}
\right).
\end{eqnarray}
A similar random term could be added to the rate of removal, however we
do not consider this problem here. Moreover, arbitrarily complex noise
inputs could be devised that would eventually reproduce features of the
observed data. Our concern is only whether these features can be
observed from the SIR model, or immediate generalisations of this model.  

We found that simulations with $\epsilon_t\sim N(0,\sigma^2)$ for a
constant $\sigma^2$ did not produce realistic simulations. The problem
is that the magnitude of the random fluctuation should be proportional to
the number of infected individuals. Few infected individuals are
unlikely to have a great variation in effect, but many such individuals
will. In Fig. \ref{sirfig} (b) we show five simulations with the noise
term $\epsilon_t\sim N(0,(I_t\sigma)^2)$. To ensure that the model is
physically realistic (i.e. non-negative populations) we also restrict
$\epsilon_t>-(rS_t+(1-a))I_t$. We found that $\sigma\approx 0.25$ gave
quantitatively reasonable results and this is the value employed in Fig.
\ref{sirfig}.

\begin{figure}[t]
\[\epsfxsize 65mm \epsfbox{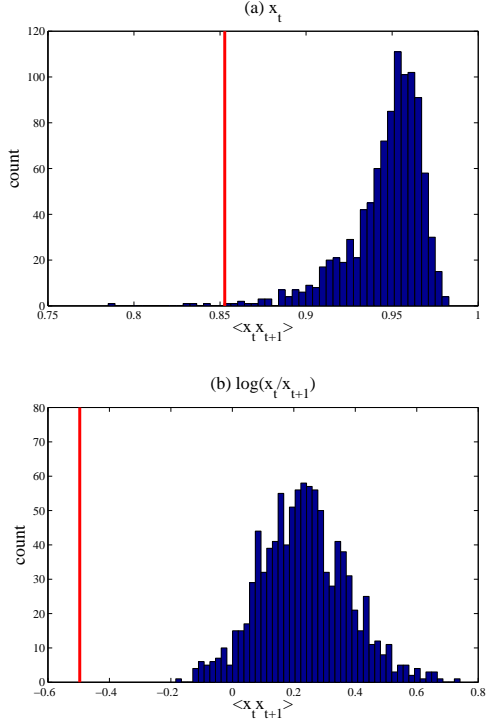}\]
\caption{{\bf Surrogate calculations for stochastic SIR data.} The normalised covariance of $x_t$ and
$x_{t+1}$, is plotted for the data along with $1000$ independent model
simulations (this is a similar calculation to that depicted in Fig. \ref{sir} (b)). The statistics in panel
(a) and (b) differ. In panel (a) the normalise covariance of the data is
estimated, in panel (b) the normalised covariance of the log-ratio of the
data is computed. In both cases the hypothesis that the SIR model is a
good model is soundly rejected ($p\ge 0.95$).}
\label{sirsurr}
\end{figure}

To quantitatively evaluate the effectiveness of the SIR type models
(Eqn. (\ref{sireqn2})) we apply a variant on the method of surrogate data
\cite{pivotal_thing}. However, instead of generating constrained
realisations consistent with the null hypothesis\footnote{See
\cite{jT96a} for a description of this rather exclusive
nomenclature. Informally this means we do not force the surrogates to
preserve specific properties of the data. This means that we are unable
to reject {\em all} SIR models with this test, only the SIR models that are
fit to this data.}, we generate multiple realisations of the model and
compare these realisations to the data \cite{fT93}. For the model
described by Eqn. (\ref{sireqn2}) we generate $1000$ simulations and for each simulation
compute $<x_tx_{t+1}>$ and $<z_tz_{t+1}>$ where
$z_t=\log\frac{x_t}{x_{t+1}}$. That is, we use two distinct test
statistics. Results for both statistics are plotted in
Fig. \ref{sirsurr} show a clear distinction between data and
surrogates. Since this model is the optimal fit to the data and the
distribution of covariance estimates do not vary greatly with change
in model parameters\footnote{That is, the statistic is approximately
pivotal \cite{pivotal_thing}.}, we therefore conclude that models of the
form Eqn. (\ref{sireqn2}) are not adequate to describe the underlying
dynamics.

\begin{figure}[t]
\[\epsfxsize 65mm \epsfbox{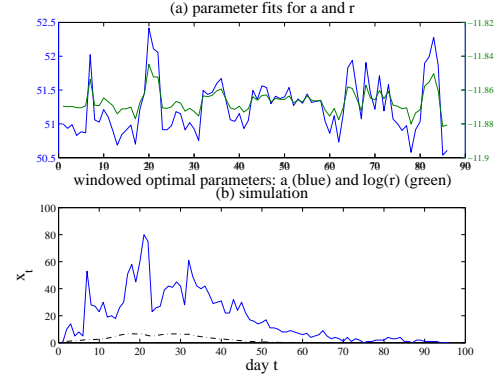}\]
\caption{{\bf Variable parameters ($a$,$r$).} The top plot shows the best fit
estimates of the parameters $a$ and $r$ where the parameters are fitted
to a window of $10$ successive data. The lower plot shows a
deterministic prediction using these variable values.}
\label{varar}
\end{figure}

One widely accepted reason for the failure of the SIR models is that the
parameters $a$ and $r$ do not remain constant during the time course of
an epidemic. Particularly when a disease is previously unknown and
public health practice changes significantly in response to the spread of
the disease. This is certainly consistent with the situation for SARS in
Hong Kong. We therefore modify the SIR model of Eqn. (\ref{sireqn}) by
estimating the temporal dependence of $a$ and $r$. We apply a sliding
window of width 10 days and re-estimate $a$ and $r$ for every day. Figure
\ref{varar} depicts the results of this computation. The values $a_t$ and
$r_t$ are based on the best fit estimate of  Eqn. (\ref{sireqn})
to the data $(x_t,x_{t+1},x_{t+2},\ldots,x_{t+10})$. 

While the simulations in Fig. \ref{varar} (b) does have the correct
shape, if not the correct magnitude, the values of $a$ and $r$ depicted
in Fig. \ref{varar} (a) illustrate the problem with this approach. The
parameter values are sensitively dependent on the observed data and do
not reflect reality. In a separate publication we modify this approach
by considering longer windows and have shown the utility of the ratio
$r/a$ for estimating the effectiveness of various governmental control
strategies \cite{sars1}. A combination of the random fluctuations approach of
Eqn. (\ref{sireqn2}) and the variability in $a$ and $r$ may produce more
realistic simulations. However, we feel that this would represent a
significant over-parameterisation of this modelling problem.

For the current investigation we conclude that a general SIR model is
unable to capture the underlying dynamics of the system. More exotic SIR
inspired models, or more contrived noise inputs, may be able to
reproduce the observed behaviour. Indeed, we expect this should be the
case when the system becomes complex enough. But this is beyond the
focus of the current communication and is in general a less interesting
problem. In the following section we consider a computationally more
extensive model that is inspired by the physical situation and provides
a more realistic alternative.

\section{Small world networks}
\label{sw}

The final model structure we examine is computationally more complex,
but also more realistic. In contrast to potentially exotic modified SIR
models, the model we introduce here is inspired by our understanding of
the physical situation. Small world networks (SWN) \cite{dW98} have recently become
immensely popular and received significant attention in a wide variety
of fields \cite{dW03}. One system which has provided evidence of
small world structure is the network of social interaction between
individuals (the so-called ``six degrees of separation''). It is
therefore natural to consider propagation of disease epidemics between
nodes of a SWN.

In the following SWN we suppose that each node is either susceptible to
disease (S), infected (I) or removed (R). However, unlike the standard
SIR structure, nodes of the SWN can only infect those that they are
acquainted with. We arrange the nodes in a rectangular grid (for Hong
Kong, we use a square with side length $2700$). On a given day, each
node can infect $n_1$ of its four immediate neighbours (horizontally and
vertically) with constant probability $p_1$. Each node also has a random number
of long range acquaintances. The number of long range acquaintances follows
an exponential distribution with an expected value of $n_2$ and the
acquaintances are chosen at random but fixed for each node. On each
day, each of the long range acquaintances may be infected with some
constant probability
$p_2$. Finally, on each day, every infected node has probability $r_1$
of becoming removed. As with the SIR model, the only possible
transitions are $S\rightarrow I\rightarrow R$. 

\begin{figure}[t]
\[\epsfxsize 65mm \epsfbox{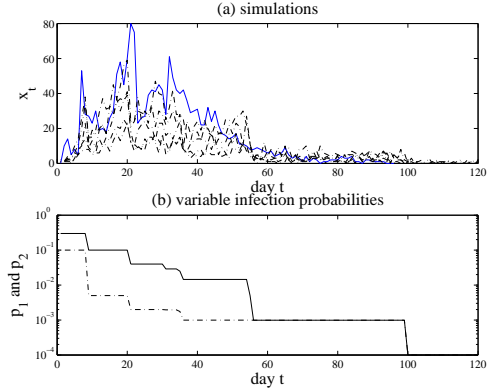}\]
\caption{{\bf Variable parameters ($p_1$,$p_2$).} The top plot is the Hong
Kong SARS data together with five simulations of small world network
model with the variables $p_1$ and $p_2$ changing with time (presumably
in response to changing public health practice). The bottom plot shows
the variation in the parameters $p_1$ and $p_2$ used in the top
plot. These values were selected using an {\em ad hoc} process.}
\label{varp12}
\end{figure}

For our first simulations we simplify this structure by setting $r_1=0$.
To avoid eventual complete contamination we vary (i.e. decrease) $p_1$
and $p_2$ with time according to the scheme in Fig. \ref{varp12}
(b). The values depicted in Fig. \ref{varp12} are selected only
because they yield a model with has a realistic behaviour. Figure
\ref{varp12} (a) compares five random realisations of $\Delta S_t$ to
the observed data. We see that the quantitative features are reproduced nicely.

\begin{figure}[t]
\[\epsfxsize 65mm \epsfbox{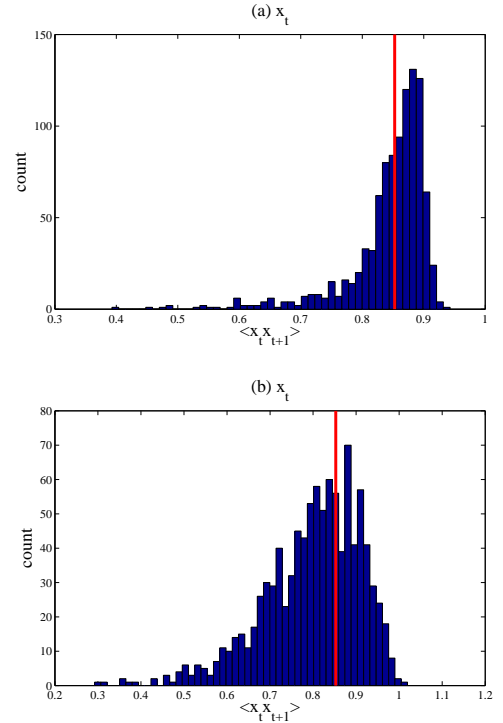}\]
\caption{{\bf Surrogate calculations.} The top plot shows $1000$
simulations of the small world network with variable $p_1$ and $p_2$
(and $r_1=0$),
the bottom plot shows estimates for $1000$ simulations from the
scale-free small world network with constant parameters ($p_1=0.05$,
$p_2=0.005$ and $r=0.10$).}
\label{smallworld}
\end{figure}

In Fig. \ref{smallworld} (a) we repeat the surrogate calculations of
the previous sections for $1000$ realisations of the process depicted in
Fig. \ref{varp12}. An examination of the variation in the estimated
covariance shows that the data and model simulations are in very close
agreement. We are therefore unable to reject the model behaviour as a
plausible model of propagation of SARS. This good agreement is due to a
combination of long range correlation present in the model data and the same
``burstiness'' that is also observed in the true data. In Fig. \ref{varp12} we see that
both model and simulations exhibit similar large spikes when one (or
several) so-called ``super-spreaders'' are encountered at random. 

However, this model is not entirely satisfactory. The disease epidemic
is controlled artificial by suitable selection of
$p_{1,2}(t)$. Furthermore, no nodes
ever become ``removed''. To remedy this we set $p_1(t)=p_1=0.05$,
$p_2(t)=p_2=0.005$, and $r_1=0.1$. These values are selected after a
process of trial and error, but do roughly correspond to the experience
of SARS in Hong Kong. The probability $p_1$ is the daily probability of
infected close family members or neighbours, $p_2$ is the daily
probability of infecting more casual acquaintances. Finally, we allow for
an exponential distribution of the number of neighbours with an
expected value of $6$. Hence the system truly is a scale free network
and the existence of ``super-spreaders'' is explicitly modelled by this
random variable. 

The time series produced by this simulation are similar to those of the
restricted SWN in Fig. \ref{varp12}. The major differences are the
inclusion of the removed category and that the
infection probabilities do not change with time. A constant  value for $p_{1,2}$ may
not be entirely realistic, but is is a statistical necessity to avoid
over-fitting in such a complex stochastic model. A consequence of these
constant values of $p_{1,2}$ is that often the disease will ``die-out''
immediately, or, it will run out of control and infect the
entire community. In simulations, both situations are common, but so to
is the eventual control of the epidemic: {\em without} changing the
infection control policy!

We do not show typical trajectories of this final model because they are
similar to those shown in Fig. \ref{varp12}, only with a slightly
larger variability. In Fig. \ref{smallworld} we generate $1000$
simulations of length greater than $90$ (i.e. simulations that
immediately die out are rejected) but no longer than $120$ (longer
simulations were truncated) of this SWN system ($n_1=4$, $E(n_2)=10$,
$p_1=0.25$, $p_2=0.001$, and $r_1=0.45$) \footnote{Simulations
were repeated with a variety of other values and a wide variety of
dynamics behaviour. Most notably the parameter values $n_1=4$, $E(n_2)=6$,
$p_1=0.05$, $p_2=0.005$, and $r_1=0.1$ produced surrogate results
similar to those in Fig. \ref{smallworld}. These two sets of parameter
values where selected as they reflected the authors' understandings of the
physical environment.} and compare the covariance of
$\Delta S_t$ to the true data. In each case we are
unable to reject the underlying SWN model. A slight increase in
variability (when compared to the previous model with artificially
variable $p_{1,2}$) is observed in this second structure and we believe that this
is due to the lower level of ``control'' on the parameter values and
the inclusion of both simulations that die out, and those that run out of control in
the analysis.

\section{Conclusion}

The daily reported SARS figures for Hong Kong have been considered. We
find that the data is not consistent with a random walk or a noisy
version of the standard SIR model. However, two SWN structures do
exhibit dynamics indistinguishable from the true data. To accurately
simulate the spread of disease in society, including features observed
during the SARS epidemic (such as rapid, extensive and localised
outbreaks), one needs to consider complex SWN type structures. The power
of the SIR model lies in its ability to model large scale spread of a
contagion, the SWN structure described here is much better at simulating
the localised variability. This localised variability, inherent to the
model simulations may allow administrators and forecasters to estimate possible variation in
their predictions of disease dynamics (i.e. to provide error bars on
their projections and evaluate the likelihood of various
scenarios). Such techniques are certain to be useful when planning and
implementing control strategies. In a related publication, we provide a
more detailed analysis of the potential scale-free structure of various
SWN models \cite{sars1}. Furthermore, in this paper we propose the
variable parameter SWN model is actually inspired by governmental
public health practice \cite{sars1}. In this current we our primary goal
is to show that SWN models are able to capture the underlying dynamics
of the SARS epidemic better than either simple stochastic or SIR models.

\section*{Acknowledgments}
This research is supported by Hong Kong University Grants Council CERG
number PolyU 5235/03E. 

\bibliographystyle{ieicetr}% bib style
%\bibliography{/home/ensmall/latex/bibliography}

%\begin{thebibliography}{99}% more than 9 --> 99 / less than 10 --> 9
%\bibitem{}
%\end{thebibliography}

%\profile{}{}
\profile*{}{}% without picture of author's face

\end{document}